\documentclass[journal=jpcafh,manuscript=article]{achemso}

\newcommand{\xp}{x^{\prime}}
\newcommand{\wfx}{\psi(x)}
\newcommand{\wfxp}{\psi(x^{\prime})}
\newcommand{\triple}{\mathrm{C}\equiv \mathrm{C}}
\newcommand{\single}{\mathrm{C}-\mathrm{C}}
\newcommand{\ch}{\mathrm{C}-\mathrm{H}}
\newcommand{\form}{\mathrm{C}_N \mathrm{H}_2}
\newcommand{\formmin}{\mathrm{C}_N \mathrm{H}_2^{-}}
\newcommand{\formo}{\mathrm{C}_N \mathrm{H}_2^{\,0}}
\newcommand{\cfourty}{\mathrm{C}_{40} \mathrm{H}_2}
\newcommand{\csixty}{\mathrm{C}_{60} \mathrm{H}_2}
\newcommand{\ceighty}{\mathrm{C}_{80} \mathrm{H}_2}
\newcommand{\chundred}{\mathrm{C}_{100} \mathrm{H}_2}

\newcommand{\dn}{\delta_n}
\newcommand{\dno}{\delta_0}
\newcommand{\qn}{q_n}
\newcommand{\rn}{\rho_n^s}
\newcommand{\qi}{q_i}
\newcommand{\ri}{\rho_i^s}
\newcommand{\Eav}{E_A^{\,\mathrm{v}}}
\newcommand{\Eas}{E_A^{\,\mathrm{s}}}
\newcommand{\Uo}{U^{0}}
\newcommand{\Um}{U^{-}}

\usepackage{graphicx,amssymb,amsmath}

\author{M.~L.~Mayo}
\author{Yu.~N.~Gartstein}
\affiliation{Department of Physics, The University of Texas at
Dallas, P. O. Box 830688, EC36, Richardson, Texas 75083, USA}
\email{yuri.gartstein@utdallas.edu}

\title{Polarons and charge carrier solvation on conjugated carbon chains: A comparative \textit{ab initio} study}

\begin{document}

\begin{abstract}
We study accommodation of an excess charge carrier on long even-$N$ polyynic oligomers $\form$ due to displacements of the underlying carbon lattice and polarization of the surrounding solvent in the context of carrier self-localization into a polaronic state. Spatial patterns of bond-length alternation, excess charge and spin densities are compared as derived with Hartree-Fock and two hybrid-DFT methods (BHandHLYP and B3LYP) in conjunction with the polarizable continuum model. Quite distinct resulting pictures of carrier accommodation are found when contributions from different interactions are analyzed. Solvation robustly acts to promote excess charge localization.
\end{abstract}

\section{Introduction}

A variety of one-dimensional (1D) semiconductor (SC) nanostructures such as $\pi$-conjugated polymers (CPs), nanowires, nanotubes, and biological macromolecules attract a great deal of attention. They are interesting scientifically and can be exploited in various technological areas including (opto) electronics, energy harvesting and sensors. The nature and properties of excess charge carriers on these structures are fundamental for many processes and applications. For purposes of our discussion here, one distinguishes between nearly free band states of excess carriers and self-localized polaronic states \cite{polarons1,appel,rashba87,alexmott}. Self-localization and the formation of 1D polarons may occur due to the (strong) interaction of the electronic subsystem with another subsystem such as displacements of the underlying atomic lattice, the mechanism  extensively studied for various 1D electron-phonon systems, and particularly for CPs \cite{YuLubook,HKSS,Barford_book}. Another, and less explored, implementation of the polaronic effect can take place for 1D SCs immersed in 3D polar media, the situation common for applications involving fundamental redox processes in polar solvents. In this case of what could be called charge carrier solvation, the long-range Coulomb interaction is expected \cite{conwell_1,YNGpol,MGShort_prb,MGLong_jcp} to lead to the formation of a localized electronic state on a SC structure surrounded by a self-consistent pattern of the sluggish (orientational) polarization of the solvent. Among consequences of self-localization,
lattice-deformation and solvation-induced polarons  can feature new signatures in the optical absorption due to local intragap electronic levels \cite{appel,YuLubook,HKSS,Barford_book,BredasPolOptics,UG_optabs} and substantially reduced mobilities \cite{appel,conwell_1,CBDrag,GU_lowfreq}.

In this paper we use three different theoretical levels of \textit{ab initio} computations to compare pictures of accommodation of an excess electron on a prototypical 1D SC system: long even-$N$ polyynic chains $\form$, emerging due to both bond-length adjustments of the carbon lattice and polarization of the surrounding solvent. The system we study is nearly charge-conjugation symmetric and similar results are obtained \cite{MGLong_jcp} upon an addition of an extra hole. In even-$N$ oligomers the arising patterns of bond-length modulations are  non-topological (polaron-like) being thus representative of a broad range of CPs.  A related comparative study of odd-$N$ polyynic oligomers has been recently published \cite{MG_kinks};  those feature topological kink-like patterns that can be found in polymers with the degenerate ground state, trans-polyacetylene being another example \cite{YuLubook,HKSS,Barford_book}.

In applying first-principles calculations one expects to be able to elucidate effects due to reorganization of valence electrons and many-electron interactions, which can be crucially important in accommodation of excess charge carriers. Numerous studies at various theoretical levels have been published on the structure of the ground state and excitations in CPs.
It is therefore noteworthy that, despite a relatively long history of electron-lattice polarons in CPs (see, e.g., multiple references in  Refs.~\cite{BredasPolOptics,bobbert}), some questions have been raised recently regarding applicability of different \textit{ab initio} frameworks to describe the polaron formation. That pertains to the reported failure of the local-density-approximation and generalized-gradient-approximation DFT (density-functional theory) schemes to detect self-localized charge density distributions in various charged oligomers, while Hartee-Fock (HF), parameterized semi-empirical and possibly hybrid-functional DFT \cite{Martin} methods have been reported to lead to charge localization in the middle of oligomers \cite{bobbert,pitea_1,bredas_3,zuppi2003}. Quite different level-of-theory-dependent magnitudes of the effective electron-phonon coupling have also been found for the dimerized ground states of polymeric systems \cite{yang_kertesz_1,yang_kertesz_2}. These observations thus bear on an important general issue of a choice of appropriate \textit{ab initio} methods \cite{Martin,StandardModel}  to faithfully describe properties of 1D semiconductors. In our computations for this paper, we compare results derived with the GAUSSIAN 03 implemented \cite{g03} HF and two popular hybrid-functional DFT methods (BHandHLYP and B3LYP) in vacuum and solvent environments, the latter represented within the framework of the polarizable continuum model \cite{leach_1,QMCSM}. We will be interested in the carbon-carbon bond-length alternation (BLA) patterns and spatial distributions of excess charge and spin densities resulting from accommodation of an extra electron.

Our choice \cite{MGShort_prb,MGLong_jcp,MG_kinks} of linear polyynic carbon chains as a model system is to a large extent related to their structural simplicity that allows us to keep computations on relatively long oligomers (up to $N=100$ in this paper) still practicable. In doing so, we attempt to address the intrinsic response of the system as determined by inherent interactions rather than by the end effects due to the hydrogen termination. In terms of the polaron formation, one would like to see if the polaron-like spatial structure is indeed self-maintained, that is, if its spatial extent is not affected by the length of the oligomer. We have explicitly demonstrated \cite{MG_kinks} in the case of kink-solitons that their resulting spatial extent strongly depends on the computational method used. Our results in this paper also display a very strong method-dependence as will be discussed later. While being a model system, one should note that polyynic chains continue to be the subject of much attention in their own right \cite{gladysz_1,gladysz_2,yang_kertesz_1,yang_kertesz_2,schaefer_1,schaefer_2,Magna09,KY2009}.
A qualitative difference in the electronic structures of polyyne and trans-polyacetylene is that the former features an extra spatial degree of freedom: $\pi$-electron molecular orbitals can have two independent orientations perpendicular to the polymer axis. Within the framework of appropriately modified Su-Schrieffer-Heeger (SSH) model of conjugated polymers \cite{YuLubook,HKSS,Barford_book}, it was shown \cite{rice_2,rice_3}  that, thanks to the extra degeneracy of $\pi$-electron levels, polyynes would
possess a particularly rich family of electron-lattice self-localized excitations. Also, spin-charge relationships for such excitations are quite different from trans-polyacetylene's.

In order to distinguish between various effects (due to electron-electron interactions, solvent polarization and carbon lattice adjustments) on resulting charge and spin densities, we study both systems with prescribed carbon atom positions (referred to as ``rigid geometry'' cases) as well as fully optimized systems, in which carbon atom positions minimize the total energy. Rigid geometries feature uniform BLA patterns chosen based on the results of optimization in the ground state of neutral oligomers $\formo$. The data on neutral oligomers is used as benchmarks against which we determine changes in BLA patterns and excess charge distributions in negatively charged oligomers $\formmin$.

\section{Computations and data processing}

All \textit{ab initio} computations in this study were performed using the
GAUSSIAN 03 suite of programs \cite{g03}. We have employed pure Hartree-Fock (HF) along
with the well-known \textit{hybrid}-DFT functionals BHandHLYP and B3LYP in order to compare
results of the different levels of theory. Hybrid exchange-correlation density functionals \cite{Martin} include both local and non-local effects and are commonly used in studies of electronic properties of CPs. The reader is reminded that functionals used here differ by the amounts of HF exchange included: $50\%$ for BHandHLYP and $20\%$ for B3LYP. A rich 6-311++G($d,p$) all electron basis set has been employed
throughout.  To investigate the effects
of solvation in a polar medium, GAUSSIAN 03 \cite{g03} offers its implementation
of the polarizable continuum model (PCM) described in original publications \cite{tomasi_1,
tomasi_2,tomasi_3,barone_1}. For all ``in solvent'' PCM calculations, water has been chosen as a polar solvent
with its default parameters in GAUSSIAN 03. In working with long oligomers we met with certain method-dependent limitations on obtaining
satisfactory levels of convergence. In order to produce reliable results
within the HF method, restricted (R) and restricted open (RO) shell wave functions were used. For the hybrid
DFT methods BHandHLYP and B3LYP fully unrestricted (U) wave functions were utilized.
Note that, within each computational method, we consistently used the same form
of the wave function.

The subject of our attention in this paper are long $\form$ chains  with even number of carbons $N$ ($N$ up to 100). It is well established  \cite{karpfen_1,yang_kertesz_1,yang_kertesz_2,pfeiffer_1,schaefer_1} that long even-$N$ neutral oligomers  have their ground state with the polyynic structure, that is, they feature an alternating pattern of triple $\triple$ and single $\single$ bonds and a gap in the electronic spectrum. We note that wherever possible, we verified that results of our ``in vacuum'' calculations compare well against previously published \cite{yang_kertesz_2,schaefer_1} \textit{ab initio} data on even-$N$ neutral and charged $\form$ systems both in terms of energetics and optimized
bond lengths. To our knowledge,  oligomer lengths explored in our studies (Refs.~\cite{MGLong_jcp,MG_kinks} and this paper) are appreciably longer than used previously in comparable calculations for charged systems.

\begin{figure}[t]
\centering
\includegraphics[scale=0.8]{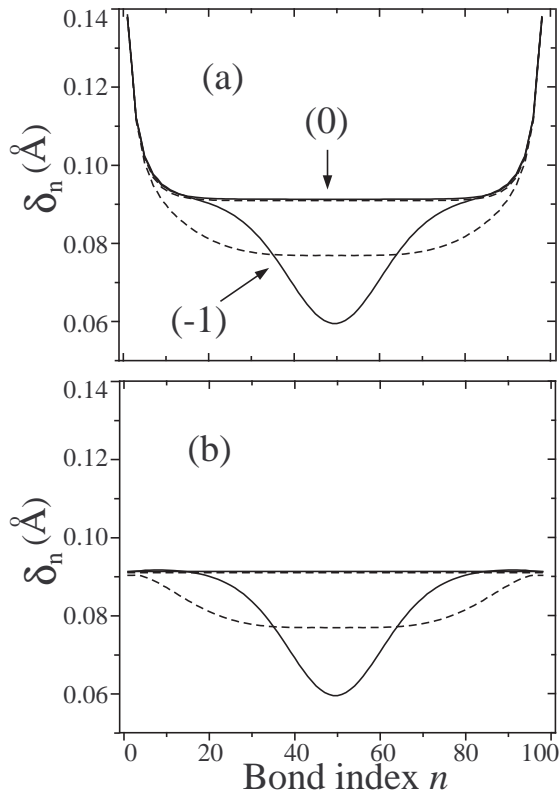}
\caption{Illustration of end effects in BLA patterns and their elimination: Panel (a)
compares BLA patterns for the neutral (0) and charged (-1) chains with end effects.
Panel (b) indicates the BLA pattern after the subtraction of common end BLA structures.
Vacuum results are shown with dashed lines and solid lines indicate the resulting
BLA pattern upon solvation. Results depicted here were obtained on the $\chundred$
chain using the B3LYP method.}
\label{figBLA}
\end{figure}

A very convenient and well-known \cite{YuLubook,HKSS,Barford_book,yang_kertesz_1,yang_kertesz_2} way to characterize the geometry of dimerized polymeric structures is via bond-length alternation (BLA) patterns
\begin{equation}\label{defd0}
\dn = (-1)^n \, \left(l_n - l_{n-1} \right),
\end{equation}
where $l_n$ is the length of the $n$th \textit{carbon-carbon} bond defined,  e.~g., as on the right of the $n$th carbon atom (we do not discuss the end bonds to hydrogens).  In the infinite dimerized neutral structure, the dimerization pattern \ref{defd0} would be uniform, that is, independent of the spatial bond position $n$, we denote the \textit{magnitude} of this equilibrium ground-state pattern as $\dno$ (this is not to be confused with a specific site value of $\delta_{n=0}$). The double degeneracy of the ground state of the infinite polymer is in that the uniform pattern \ref{defd0} can assume either value of $+\dno$ or $-\dno$ \cite{YuLubook,HKSS,Barford_book}. For our purposes, we retrieve the magnitude $\dno$ from the middle segment of long even-$N$ neutral oligomers featuring a well-established spatially-independent behavior, see illustration in \ref{figBLA}, and display that ground-state BLA pattern as positive. As mentioned earlier, the magnitude $\dno$ is found to be strongly method-dependent; this dependence is reflected in results shown in \ref{RG_OPT} and in published \cite{yang_kertesz_2,MG_kinks} tabulated numerical data.

While exhibiting a spatially-independent behavior in the middle of long neutral oligomers, \ref{figBLA}(a) also clearly shows that the end (oligomer termination) effects on the BLA pattern are quite substantial. In numerous computations we observed that the ``shape'' of the end-specific BLA structures stabilizes and becomes easily discernible in long enough oligomers (of course, within a specific computational method). Practically the same persistent shape has been found in end BLAs of neutral and charged, even- and odd-$N$ oligomers \cite{MGLong_jcp,MG_kinks}. This fact allows to ``eliminate'' the end effects by comparison and an appropriate ``subtraction'' procedure, as illustrated in  \ref{figBLA}(b). Such a procedure of removing the end effects from the finite oligomer data has been used to produced the BLA patterns shown in \ref{RG_OPT} focussing thus on the changes induced by the accommodation of an excess charge carrier.

Very essential for our discussion are the resulting \textit{spatial charge and spin distributions} over the oligomers, which are naturally related to atomic charges and spins in outputs of \textit{ab initio} calculations. Importantly, these quantities are calculated from many-electron wave functions and thereby reflect responses due to \textit{all} electrons in the system. Two procedures, after Mulliken and L\"{o}wdin, are widely used for charge population analysis \cite{leach_1} and only Mulliken procedure is available for spins in GAUSSIAN 03. It is well known \cite{leach_1} that calculations of atomic-centric quantities depend on the basis set transformations and sometimes lead to artifacts and spurious results. We have demonstrated previously \cite{MGLong_jcp} that  certain cancelation effects take place and Mulliken and L\"{o}wdin procedures lead to more similar results when one is interested in the spatial distribution of the \textit{excess} charge derived as the difference of charge densities on charged and neutral chains. An example of L\"{o}wdin-derived excess atomic charge densities $\qn$ from our current computations is shown in \ref{figRAW}(a). \ref{figRAW}(b) illustrates an instance of Mulliken-derived atomic spin densities $\rn$. When used for absolute data values, we keep both densities normalized to unity:
\begin{figure}[t]
\centering
\includegraphics[scale=0.8]{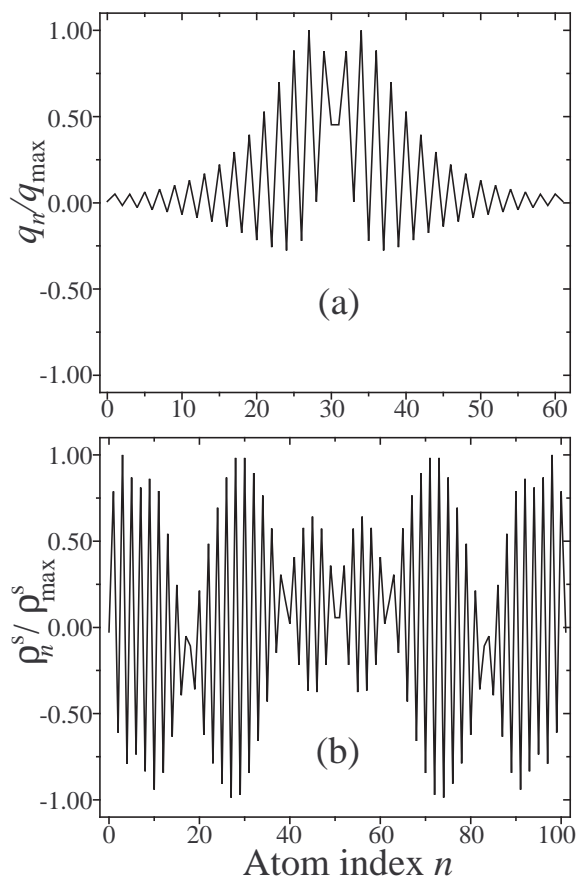}
\caption{Illustrations of the spatial behavior of raw \textit{atomic} densities scaled with respect to their maximum values. Both examples are derived from oligomers with uniform rigid geometries. (a) Excess charge density obtained for $\csixty^{-}$ in vacuum with the HF method. (b) Spin density obtained for $\chundred^{-}$ in solvent with the BHandHLYP method.}
\label{figRAW}
\end{figure}
$$
\sum_n \qn = \sum_n \rn = 1.
$$
In \ref{figRAW}, the data is shown in a scaled form to emphasize its spatial behavior. It is evident from this figure that atomic-centric quantities generally exhibit very strong variations from carbon to the next carbon. It is not clear to what extent these oscillations may result from computational artifacts perhaps also related to the presence of bond-oriented density waves \cite{YuLubook,HKSS,Barford_book} in our system. In order to eliminate (possibly spurious) high spatial-frequency effects and to focus on longer spatial-range structures, some averaging/smoothing of the atomic data can be applied. Needless to say that results can thus depend on details of the smoothing procedure. In our earlier work \cite{MGLong_jcp}, we have shown that a simple cell averaging of excess charge data can already lead to quite smooth results, moreover it has made Mulliken- and L\"{o}wdin-derived results nearly identical. The unit cells here are defined as consisting of pairs of neighboring carbon atoms starting from the oligomer ends (as is common, the chain end cells also include hydrogen atomic charges/spins). We apply the same summation over the neighboring carbons to generate cell-centric data in this paper. To smooth data even further, we also apply a simple-minded nearest-neighbor-cell averaging where a cell-specific quantity $x_i$ is transformed into $\langle x_i \rangle = 0.5 x_i + 0.25 (x_{i-1} + x_{i+1})$. So generated smoothed data on cell-centric charge $\qi$ and spin $\ri$ densities are used in \ref{RG_OPT}. While the number of cells (index $i$) is of course twice as small as the number of carbons, the procedure preserves the normalization:
$$
\sum_i \qi = \sum_i \ri = 1.
$$
While reiterating the caveat on the procedure dependence (especially on some spin data), we mention here that the described above averaging produce results with smaller-amplitude variations than some other approaches we tried, which is a desirable feature in the absence of other criteria to ascertain the physical validity of the atomic-centric data from computations.

As discussed in the Introduction, in addition to the fully relaxed geometries of the underlying atomic lattice that minimize the total system energy, we also study rigid geometries (RG) with prescribed bond lengths. Those have been generated separately for each of the computational methods used in respective environments (vacuum and solvent). We would first find a fully optimized geometry of the $N=100$ neutral oligomer and retrieve the resulting lengths of the $\triple$ and $\single$  bonds
in its central part (as well as the length of the terminating $\ch$ bonds). We then assign those bond lengths alternatingly to all neighboring carbon pairs throughout the oligomer. The resulting RG structures thus exhibit entirely spatially uniform carbon-carbon dimerization patterns within the confines of terminating hydrogens. We note that there is very little difference, as clearly seen in \ref{figBLA}, in the dimerization amplitude $\dno$ between neutral chains in vacuum and in solvent.

\section{Results and discussion}

\begin{figure}[!t]
\centering
\includegraphics[scale=0.47]{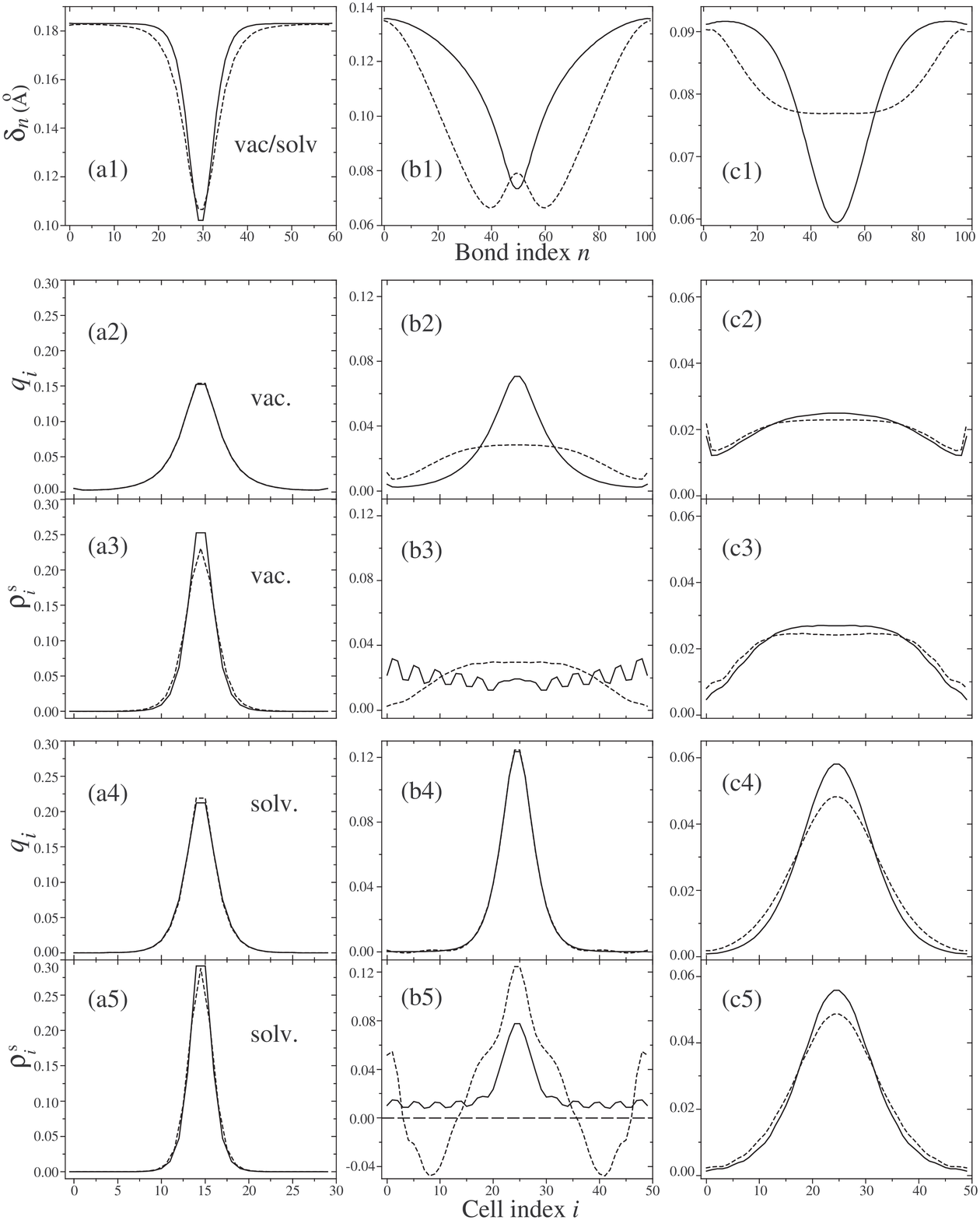}
\caption{Processed results for charged $\formmin$ oligomers as arranged in columns by the computational method: (a) ROHF, (b) UBHandHLYP, and (c) UB3LYP. Row 1 displays optimized BLA patterns with end effects eliminated for oligomers in vacuum (dashed lines) and in solvent (solid). The notation is different in rows 2--5, where solid lines show results for fully optimized carbon lattice geometries, while dashed lines (when distinguishable) for rigid geometries with uniform dimerization patterns (see the text). These rows display cell-centric excess charge and spin densities for chains in vacuum (rows 2 and 3) and in the solvent environment (rows 4 and 5).}
\label{RG_OPT}
\end{figure}

In this Section we discuss our computational results processed as described above. Our discussion will revolve around \ref{RG_OPT} displaying spatial distributions of BLA patterns, excess charge and spin densities for $N=60$ (ROHF computations) and $N=100$ (UBHandHLYP and UB3LYP) polyynic oligomers $\formmin$ with an extra electron.

It is instructive to first recall how self-trapping of an excess charge carrier (an electron or a hole) is described in a \textit{single (quasi)particle} picture within a standard 1D continuum adiabatic framework \cite{rashba87}. The self-trapped state corresponds to the localized ground-state wave function $\wfx$ in the solution of the non-linear Schr\"{o}dinger equation:
\begin{equation}\label{NSE}
-\frac{\hbar^2}{2m}\frac{\partial^2 \wfx}{\partial x^2} \,-\,
\int \! d\xp V(x-\xp)\, |\wfxp|^2 \,\wfx = E\wfx,
\end{equation}
where $x$ is the coordinate along the structure axis, $m$ the intrinsic effective mass of the carrier, and $V(x)$ the effective \textit{self-interaction} mediated by another subsystem. In the case of a short-range electron-phonon mediation, the self-interaction can be taken local: $V(x)=g\,\delta(x)$ leading then to the widely known exact result \cite{rashba_LP,holstein_LP} $\wfx \propto 1/\cosh(gmx/2\hbar^2)$ for a continuum 1D polaron.  In the case of the long-range polarization interaction, the effective $V(x)$ behaves as $1/|x|$ at large distances, while the short-range behavior depends on specific system details such as, e.g., the actual transverse charge density distribution and the geometry of the dielectric screening \cite{YNGpol,UG_optabs}. In the absence of the self-interaction, \ref{NSE} would just yield spatially delocalized plane-wave solutions. Among features of this single-particle description is that the same probability density $|\wfx|^2$ determines both spatial distributions of charge and spin of the charge carrier. The other one is that it involves the adiabatic framework, that is, the mediating subsystem is assumed very slow allowing to determine electronic states for its given static configuration, which in turn self-consistently responds to the probability $|\wfx|^2$ (``strong-coupling polarons'' \cite{polarons1,appel,rashba87}). It should be emphasized that in realizations of the computational methods used, both displacements of the underlying carbon lattice and solvent polarization constitute such slow subsystems.

The single-particle description has the drawback of not \textit{explicitly} including  valence band electrons, whose reorganization may be important in the accommodation of the excess carrier. Such a reorganization can take place even in electron-lattice models without direct electron-electron interactions that have been used to describe non-linear localized excitations in CPs \cite{YuLubook,HKSS}. A nice analysis \cite{FVB} was in fact given of how polarons of a two-band Peierls dielectric model of CPs evolve into single-particle Holstein polarons in the limit of the ``frozen valence band'' approximation. Polaronic and solitonic excitations in these models are also derived within self-consistent adiabatic treatment of the underlying lattice. In \textit{ab initio} computations, reorganization of valence band electrons occur as driven by both the interaction with the bond-length modulations and by electron-electron Coulomb interactions. With many electrons determining the resulting spatial densities, excess charge and spin distributions should not be necessarily the same. One of the famous instances of this phenomenon is inverse charge-spin relationships for kink-solitons in trans-polyacetylene \cite{YuLubook,HKSS} (these relationships are different for kinks in polyynes \cite{rice_2,rice_3}). We will also mention in this regard separation of charge and spin degrees of freedoms well known in 1D electronic liquids \cite{Gia2004,GV2005}. Still one could expect that, for non-topological localized polaronic excitations, excess charge and spin density distributions would closely follow each other.

Turning now to results in \ref{RG_OPT}, we note upfront that, similarly to our observations on kink-solitons \cite{MG_kinks}, spatial extents of the structures developed as a result of electron accommodation are strongly computational method-dependent. The magnitudes $\dno$ of the uniform dimerization are consistent with the notion of the increased effective electron-lattice interactions with the increase of the amount of the HF exchange in the method functionals. We would be more interested now in qualitative aspects of the results as related to the phenomenon of self-localization.

The main reason for our studying rigid geometry (RG) cases has been to try to distinguish effects due to different interactions. As RGs feature spatially uniform BLA backgrounds, one might expect that \textit{in vacuum} the excess electron would be delocalized over the oligomer showing the particle-in-the-box type behavior for charge and spin densities. On the other hand, if the lattice is allowed to relax, that could cause, as per a conventional adiabatic polaron picture, spatial localization of excess charge and spin. Vacuum results for the densities are shown in Rows 2 and 3 of \ref{RG_OPT}. One indeed observes that RG data displays a delocalized behavior for both charge and spin when calculated with hybrid-DFT methods. Not so however when calculated in HF. Hartree-Fock data (panels (a2) and (a3)) display localization of excess charge and spin already in RG, and the degree of localization is only very weakly affected by a quite substantial dip in the optimized BLA pattern (panel (a1)) when the lattice is allowed to relax. We conclude that self-localization observed in these results is \textit{not caused} by the lattice deformation but rather by electron-electron interactions as treated in HF. The deformation of the BLA pattern then results as a response to the developed localized density. We easily confirmed this kind of localization due to electron-electron interactions alone in simplified calculations once these interactions are treated in the HF approximation. One cannot help wondering if this qualitative picture may be just an artifact of the HF treatment.  It is not clear if there are slow (low-frequency) electronic excitations in this system that exhibits a well-known very substantial HF energy gap \cite{yang_kertesz_2,MG_kinks}. An alternative to that picture would be that
as a result of Coulomb interactions with the fast
electronic subsystem, the excess charge carrier remains delocalized but with its properties renormalized (see, e.g., calculations of corrections to the DFT quasi-particle energies within the framework of the GW-approximation \cite{Martin,StandardModel,GV2005}). Such a renormalized quasi-particle could then interact with the lattice displacements to form an electron-lattice polaron. Perhaps a properly realized time-dependent computational scheme could clarify the question of the purely electronic response.

While yielding a delocalized behavior in RG, hybrid-DFT computations exhibit other issues in the context of the polaron formation due to carbon lattice displacements in vacuum. Particularly, as has already been noticed \cite{MGLong_jcp}, we do not observe electron self-localization in results obtained with UB3LYP -- the slight changes in densities due to relaxation (panels (c2) and (c3)) and a flat optimized BLA pattern in the middle of oligomer (panel (c1)) indicate that they are finite oligomer-size effects -- at least within the range of oligomer lengths studied. With UBHandHLYP data, visually one could probably identify the excess charge density (panel (b2)) in the optimized system as localized. There is however no counterpart to this behavior in the spin density (panel (b3)), which rather looks like a spin-density wave (SDW). Also quite unusual is the shape of the dip in the optimized BLA pattern in panel (b1), quite substantial in magnitude but the spatial extent of which is still likely limited by the oligomer ends rather than being self-consistent.

Concluding the survey of the vacuum data in rows 2 and 3, \textit{none} of the methods we used seems to have produced results that can be qualified with certainty as self-localization of the excess electron charge \textit{and} spin densities \textit{caused} by its interaction with the displacements of the underlying carbon lattice.

Quite substantial effects take place due to the polarization of the surrounding solvent by the excess electron. They are easily discernible in \ref{RG_OPT}. One immediately notices in row 1 that optimized BLA patterns in the solvent acquire a (much) more spatially localized character. Results for the charge and spin densities in the solvent environment are shown in rows 4 and 5.

Using B3LYP computations, we have previously demonstrated \cite{MGLong_jcp} that the solvation effect \textit{alone} can lead to self-localization of excess charge densities. RG data in panels (c4) and (c5) clearly show that this is the case for both charge and spin densities, even though a complete self-consistency is not completely achieved yet at this oligomer length. The behavior of spin and charge distributions here closely follow each other. As HF computations yield localized patterns already in RG, one cannot study the solvation effect here separately. It is however evident from panels in column (a) that solvation results in increased spatial localization of excess charge and spin (all HF data exhibit a noticeably larger degree of spin localization than that of charge). The BHandHLYP data in column (b) also shows a significant degree of localization of charge density due to solvation (b4), appreciably more than what is caused by the lattice deformation in vacuum (b2). Once again, though, the spin response (b5) in these computations for RG is quite surprising: while, perhaps, exhibiting a localized pattern, it is accompanied by what appears to be a large amplitude and spatial scale SDW-like response. One, in fact, would not easily discern a localized pattern in the raw atomic data shown in \ref{figRAW}(b), and different averaging procedures could lead to different conclusions. The spin density distribution in this case however does acquire a more delineated localized shape, even if on the background of a smaller SDW-like pattern, when the carbon lattice is allowed to relax. The lattice relaxation does not appear to affect the already well-localized excess charge density in panel (b4). Very small, and apparently in different directions for charge and spin, is the lattice relaxation effect in the HF data ((a4) and (a5)). The only computational method that clearly shows a synergistic amplifying localization effect of solvation and lattice relaxation on both spin and charge densities is B3LYP ((c4) and (c5)). One can however say that when solvation and lattice relaxation act together, they do cause the localization of excess charge and spin densities in BHandHLYP computations as well (as compared to the delocalized behavior in the vacuum RG environment). Overall, it is evident that, independently of the computational method, solvation always acts as to boost the degree of localization of excess charge density -- the result quite understandable on simple physical grounds. It is hard to say if the peculiarities of the spin responses (and perhaps related BLA patterns) found in BHandHLYP results are caused by some computational artifacts or may indeed have a physical origin in different behaviors of charge and spin degrees of freedom due to many-electron interactions. It would be interesting to see if similar results are obtained in long C$_{4n+2}$H$_2^{-}$ oligomers that have a different electronic symmetry \cite{schaefer_1} than C$_{4n}$H$_2^{-}$ systems studied in this paper.

A truly self-consistent localized solution should be (practically) oligomer-length independent, which could be straightforwardly verified by comparing all spatial patterns (BLA, charge and spin density distributions) on oligomers of increasing lengths. That is, for instance, the case when we compare relevant results obtained in the HF computations for $N=40$ and $N=60$ oligomers. On the other hand, vacuum BLA patterns obtained in hybrid-DFT computations appear to continue to evolve even at our current limit of $N=100$ oligomers. The difference between results derived for oligomers of different lengths is substantially much smaller for computations in the solvent environment. This is in agreement with the observation evident from \ref{RG_OPT} that solvation generally leads to more spatially localized patterns, which are therefore much weaker affected by chain ends in our longest oligomers.

One can also look at the self-consistency issue from the viewpoint of energetics. \ref{tbl2} lists some energy parameters derived from the computational outputs for oligomers of different lengths.
A well-defined and physically important quantity characterizing accommodation of an excess electron is the electron affinity $E_A$,
corresponding to the difference of total system energies for neutral $\form$ and negatively charged $\formmin$ chains. \ref{tbl2} features electron affinities $\Eav$ in vacuum and $\Eas$ in solvent environments, all obtained for fully optimized systems. Needless to note that numerical values in the table are strongly method-dependent. One can easily see that, within each method, the values of $\Eas$ change only little when going to longer oligomers, an indication of the convergence to self-consistent solutions. Vacuum values $\Eav$ change by larger quantities, which is contributed to by two factors: vacuum structures are farther away from self-consistency, and the unscreened Coulomb field of the excess charge is long-range (long-range electron polarization effects may be reflected in the illustration of \ref{figRAW}(a), see also Ref.~\cite{MG_kinks}). The solvent environment efficiently screens such long-range Coulomb effects \cite{MG_kinks}.

\begin{table}
\caption{\label{tbl2}Electron affinities and polarization energies in eV}
\begin{tabular}{l | c | c c c c }
\hline
\hline
method & system &  $\Eav$ & $\Eas$ & $\Uo$ & $\Um$ \\
\hline
 & $\cfourty$ & 0.88 & 2.10 & 0.38 & 1.84 \\
\raisebox{1.5ex}[0cm][0cm]{ROHF} & $\csixty$ & 0.92 & 2.11 & 0.44 & 1.87\\
\hline
& $\csixty$ & 3.00 & 3.82 & 0.34 & 1.39 \\
\raisebox{1.5ex}[0cm][0cm]{UBHandHLYP} & $\chundred$ & 3.11 & 3.85 & 0.40 & 1.40 \\
\hline
& $\ceighty$ & 3.92 & 4.41 & 0.30 & 0.94 \\
\raisebox{1.5ex}[0cm][0cm]{UB3LYP} & $\chundred$ & 4.00 & 4.42 & 0.32 & 0.92\\
\hline
\hline
\end{tabular}
\end{table}

\ref{tbl2} also shows electrostatic energies $\Uo$ and $\Um$ stored in the solvent polarization for, respectively, neutral and charged oligomers, as retrieved from GAUSSIAN 03 outputs. We previously \cite{MGLong_jcp} analyzed the evolution of these quantities with oligomer lengths and found that $\Uo$ grows linearly with the length while $\Um$ first decreases with the length, achieves its minimum when the length becomes larger than the self-consistent size of the self-localized charge carrier, and then continues to increase just as $\Uo$ does. By these features, one can also judge from the table data that self-consistency has been practically achieved in ROHF and UBHandHLYP computations but longer oligomers would need to be employed to observe explicit self-consistency in UB3LYP computations.

Looking at the trends displayed in results obtained with two hybrid-DFT methods, we cannot exclude that there may be a range of ``customized'' hybrid functionals with the ``right'' mixture of HF and DFT correlations that would yield results featuring self-localization behavior for an excess charge carrier consistent with the single-particle picture of \ref{NSE}. One could then observe that in the absence of the interaction with a slow subsystem (uniform RG in vacuum) the resulting charge and spin densities are delocalized, while self-localization  takes place due to the lattice deformation and due to solvation as separate effects as well as synergistically. In addition, the resulting localized charge and spin densities would largely follow each other even if with small variations due to many-electron interactions. Another conjecture also suggested by our data is that the failure to detect the formation of electron-lattice polarons in pure DFT and B3LYP computations may be related to the unscreened long-range Coulomb field of the excess charge in vacuum. Indeed, we clearly detected the formation of lattice kink-solitons in such computations \cite{MG_kinks}, and electron-lattice polarons can be thought of as bound states of kink-antikink pairs \cite{YuLubook,HKSS}. The screening of long-range Coulomb forces by fast polarization of the surrounding medium could energetically stabilize the polaron formation. Fast polarization here would act only as to reduce the magnitude of Coulomb forces. This screening effect is different from nearly static reorganization of a slow solvent considered in this paper and demonstrated to be a robust promoter of excess charge localization. Exploring the above conjectures could be an interesting avenue for further developments.

\bibliography{abinitio_ref}

\end{document}